# Topological Hall Effect in a Topological Insulator Interfaced with a Magnetic Insulator

Peng Li[1]*, Jinjun Ding[1]*, Steven S.-L. Zhang[2,3]*, James Kally,[4] Timothy Pillsbury[4], Olle G. Heinonen[2], Gaurab Rimal[5], Chong Bi[6], August DeMann[1], Stuart B. Field[1], Weigang Wang[6], Jinke Tang[5], J. S. Jiang[2], Axel Hoffmann[2,7], Nitin Samarth[4], and Mingzhong Wu[1]†

[1]Department of Physics, Colorado State University, Fort Collins, Colorado 80523, USA
[2]Materials Science Division, Argonne National Laboratory, Lemont, IL, 60439, USA
[3]Department of Physics, Case Western Reserve University, Cleveland, OH, 44106, USA
[4]Department of Physics, Pennsylvania State University, University Park, Pennsylvania 16802, USA
[5]Department of Physics & Astronomy, University of Wyoming, Laramie, WY 82071, USA
[6]Department of Physics, University of Arizona, Tucson, AZ 85721, USA
[7]Department of Materials Science and Engineering, University of Illinois, Urbana, IL 61801, USA

**ABSTRACT:** A topological insulator (TI) interfaced with a magnetic insulator (MI) may host an anomalous Hall effect (AHE), a quantum AHE, and a topological Hall effect (THE). Recent studies, however, suggest that coexisting magnetic phases in TI/MI heterostructures may result in an AHE-associated response that resembles a THE but in fact is not. This article reports a genuine THE in a TI/MI structure that has only one magnetic phase. The structure shows a THE in the temperature range of $T$=2-3 K and an AHE at $T$=80-300 K. Over $T$=3-80 K, the two effects coexist but show opposite temperature dependencies. Control measurements, calculations, and simulations together suggest that the observed THE originates from skyrmions, rather than the coexistence of two AHE responses. The skyrmions are formed due to a Dzyaloshinskii–Moriya interaction (DMI) at the interface; the DMI strength estimated is substantially higher than that in heavy metal-based systems.

A topological insulator (TI) is electrically insulating in its interior but hosts conducting states on its surfaces. The topology of the surface states in a TI is protected by time-reversal symmetry; in momentum space, such topological surface states (TSS) manifest themselves as a Dirac cone with spin-momentum locking. One can break the time-reversal symmetry of the TSS and thereby open a gap at the Dirac point by interfacing the TI with a magnetic insulator (MI) with perpendicular magnetization. Such manipulation of the TSS can give rise to exotic quantum effects. In fact, previous experiments have already demonstrated the quantum anomalous Hall effect (QAHE) and axion insulator states in the region where the Fermi level ($E_F$) is within the magnetic gap opened by the MI,[1,2,3] and the anomalous Hall effect (AHE) in the case where $E_F$ is not located in the magnetic gap.[4,5,6,7,8] Since these effects have fundamental and technological implications, they have attracted considerable interest recently.

In the region where $E_F$ is not in the magnetic gap, a topological Hall effect (THE) has also been reported, adding a new member to the "Hall" family of the TI.[9,10] The experiments were carried out by interfacing a TI film with a magnetic TI film[9] or sandwiching a TI film with two magnetic TI layers,[10] and the THE was measured as an "excess" Hall signal on top of the AHE signal. In contrast to the AHE and the QAHE that arise from the Berry curvature in momentum space,[4,5,11] the THE originates from the Berry curvature of topological spin textures in real space.[12,13,14,15,16]

However, a very recent experiment[17] offered an alternative interpretation of the THE reported in TI heterostructures[9,10] — it attributes the THE signals to the overlapping of two AHE signals with opposite signs. A similar interpretation has also been offered to explain THE-like signatures in SrTiO$_3$/SrRuO$_3$/SrTiO$_3$ heterostructures.[18] In such a scenario, the competing AHE signals that produce a THE-like signal could arise either from the coexisting surface and bulk magnetic phases in the magnetic TI layer, or two interfaces in the MI/TI/MI tri-layer case. However, in a completely TI-based heterostructure where the magnetic properties and the current flow paths are complex, it may be difficult to definitively distinguish a genuine THE associated with real-space topological spin texture from the competing AHE scenario. It is thus important to study THE in a structure that simplifies the sample geometry.

This article reports THE responses in a simple bi-layered structure that consists of a TI Bi$_2$Se$_3$ thin film grown on top of a MI BaFe$_{12}$O$_{19}$ thin film. As opposed to previous TI structures,[9,10,17] this bi-layer facilitates the demonstration of *bona fide* THE phenomena in TIs for two reasons: (1) the BaFe$_{12}$O$_{19}$ film is a well-characterized, insulating, single-phase magnetic material; (2) only one of the Bi$_2$Se$_3$ film surfaces is interfaced with a MI. These facts exclude the possibility of the coexistence of two AHE signals[17] and thereby establish the Bi$_2$Se$_3$/BaFe$_{12}$O$_{19}$ bilayer as a clean, model system for THE studies.

The magnetization in BaFe$_{12}$O$_{19}$ is perpendicular to the film plane, thanks to magneto-crystalline anisotropy.[19] $E_F$ in Bi$_2$Se$_3$ is in the bulk conduction band,[6] not in the gap opened by BaFe$_{12}$O$_{19}$, setting the bilayer to the AHE regime rather than the QAHE regime. The structure showed pure AHE responses over a temperature ($T$) range of 80–300 K, but pure THE responses at $T$ = 2–3 K. Over $T$ = 3–80 K, the two effects coexist, but exhibit opposite $T$ dependences. The procedure in Ref. [17] is used to attempt to fit the THE signals with two AHE components. While the fits are very good for the THE data at each temperature, the temperature dependences of the extracted two potential AHE contributions are completely unphysical, confirming that the observed THE does not result from two overlapping AHE signals.

Control measurements and theoretical and numerical analyses indicate that the observed THE originates from the presence of a Dzyaloshinskii–Moriya interaction (DMI) at the interface due to strong spin-orbit coupling in Bi$_2$Se$_3$ and broken spatial inversion symmetry of the bi-layer. The DMI induces skyrmions in BaFe$_{12}$O$_{19}$, while the latter give rise to the THE in Bi$_2$Se$_3$ via spin-dependent scattering at the interface. The $T$ dependence of the THE signal is related to that of the DMI strength and the carrier density, while the $T$ dependence of the AHE is associated with that of the conductivity of the TSS and the phonon density.

These results have significant implications. From a fundamental point of view, they demonstrate, for the first time, a genuine THE in TI materials; there have been two reports on THE responses in TIs,[9,10] but they may arise from the coexistence of two magnetic phases.[17] Our results are also of great technological interest in view of potential applications of skyrmions as information carriers. First, they demonstrate a new approach for skyrmion realization that uses a robust TI/MI bilayer in a geometry that can be appropriately engineered to work at technologically relevant temperatures. As discussed shortly, the spin-momentum locking of the TSS enables the presence of an interfacial DMI that is substantially stronger than in heavy metal-based structures; strong DMI is essential for the realization of small-size skyrmions. Second, the insulating nature of MI films, in general, precludes direct, electrical detection of chiral spin textures in the films; this work demonstrates the feasibility of electrical reading of skyrmions in a MI film by a TI film, thus providing a route toward proof-of-concept demonstrations of MI-based skyrmion devices.

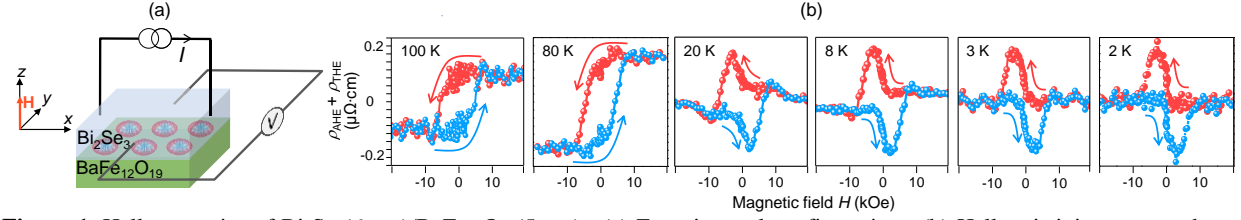

**Figure 1.** Hall properties of Bi$_2$Se$_3$(6 nm)/BaFe$_{12}$O$_{19}$(5 nm). (a) Experimental configuration. (b) Hall resistivity measured as a function of a perpendicular field ($H$) at different temperatures.

Figure 1 presents the main data of this work. They were obtained on a Hall bar structure made of a 6-nm-thick Bi$_2$Se$_3$ film grown on a 5-nm-thick BaFe$_{12}$O$_{19}$ film. Details about the material growth and characterization are presented in Supplemental Material S1-S4. In brief, the BaFe$_{12}$O$_{19}$ film was grown by pulsed laser deposition, the Bi$_2$Se$_3$ film was grown by molecular beam epitaxy, and the Hall bar was fabricated through photolithography and ion milling processes.

Figure 1(a) illustrates the experimental configuration, while Fig. 1(b) gives Hall resistivity ($\rho_H$) vs. field ($H$, perpendicular) responses measured at six different temperatures ($T$). In general, $\rho_H$ can be expressed as[12,13]

$$\rho_H = \rho_{OHE} + \rho_{AHE} + \rho_{THE} = R_0 H + R_a M + \rho_{THE} \quad (1)$$

where $\rho_{OHE}$, $\rho_{AHE}$, and $\rho_{THE}$ denote the ordinary Hall effect (OHE), AHE, and THE resistivities, respectively, $R_0$ and $R_a$ are constants, and $M$ is the magnetization. For the data in Fig. 1(b), $\rho_{OHE}$, scaling linearly with $H$, has already been subtracted, and the vertical axis shows $\rho_{AHE} + \rho_{THE}$. At $T=100$ K, the data show a nearly-square loop that is similar to the magnetic hysteresis loop of BaFe$_{12}$O$_{19}$ [see Fig. S1] and is typical for the AHE.[4-7] With a decrease in $T$, however, a hump-like structure develops on the top of the loop, indicating the coexistence of the AHE and the THE.[12-16] As $T$ is further decreased to 3 K and then to 2 K, the square loop response disappears, and one observes only peak- and dip-like responses expected for the THE,[12-16] indicating the existence of a pure THE.

The THE responses in Fig. 1(b) cannot be the result of two overlapping AHE responses because (i) the TI/MI structure concerned here has only one magnetic phase (BaFe$_{12}$O$_{19}$, insulating), and the two magnetic sublattices in BaFe$_{12}$O$_{19}$ do not give two AHE signals (Supplemental Material S10); (ii) the structure has only one conducting layer (Bi$_2$Se$_3$); and (iii) the structure has only one TI/MI interface, and the TI top surface is non-magnetic and does not produce AHE signals. Thus, the TI/MI structure is expected to show only one AHE, in stark contrast to the previous work.[9,10,17] To confirm this, the data in Fig. 1(b) were fitted to[17]

$$\rho_{AHE}(H) = \rho_{AHE1} \tanh\left(\frac{H \pm H_{c1}}{H_{01}}\right) + \rho_{AHE2} \tanh\left(\frac{H \pm H_{c2}}{H_{02}}\right) \quad (2)$$

where the first and second terms on the right side denote the two possible AHE contributions: AHE1 and AHE2. $\rho_{AHE1}$ ($\rho_{AHE2}$) and $H_{c1}$ ($H_{c2}$) are the amplitude and coercivity, respectively, of the resistivity vs. field loop of AHE1 (AHE2); $H_{01}$ and $H_{01}$ are two constants.

The fitting results are presented in Fig. 2. The curves in Fig. 2(a) present the fits of the data measured at $T=2$ K, while the corresponding AHE1 and AHE2 components of the fitting are given in Fig. 2(b). Note that the AHE2 loop is too narrow to be visible. A similar fitting was carried out to the data measured at different temperatures. The $\rho_{AHE1}$ and $\rho_{AHE2}$ values obtained from the fits are presented in Fig. 2(c), while the extracted $H_{c1}$ and $H_{c2}$ values from the fits are in Fig. 2(d). One can see that the fitting in Fig. 2(a) is almost perfect, but none of the four responses in Figs. 2(c) and 2(d) show meaningful or physical temperature dependences. Note that the resistivity of $Bi_2Se_3$ does show a typical temperature dependence [see Fig. S2]. These observations evidently show that the temperature dependences of the extracted two AHE contributions are completely unphysical, and the observed THE cannot be attributed to the superposition of two AHE signals with opposite signs. Fitting with Langevin functions was also performed, and the same results were obtained (Supplemental Material S9).

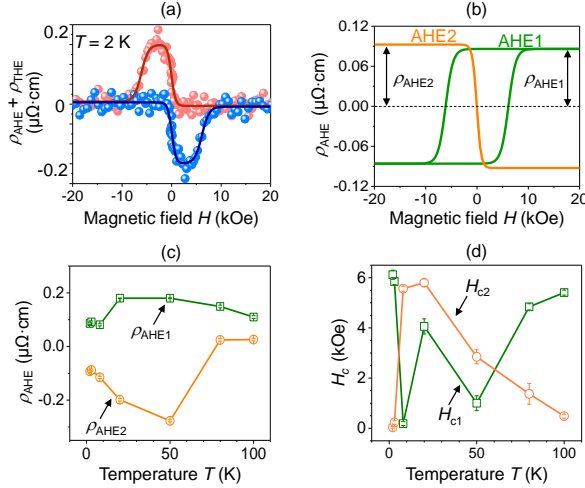

**Figure 2.** Fitting of THE data on a $Bi_2Se_3/BaFe_{12}O_{19}$ bi-layered structure with two distinct AHE contributions. The fits (curves) in (a) include two components: AHE1 and AHE2, which are shown in (b). (c) and (d) present the fitting-yielded amplitude ($\rho_{AHE}$) and coercivity ($H_c$) values, respectively, of the AHE1 and AHE2 resistivity vs. field loops.

Instead, the observed THE can be attributed to the presence of skyrmions created by the DMI at the $Bi_2Se_3/BaFe_{12}O_{19}$ interface. There is no bulk DMI in $BaFe_{12}O_{19}$ due to the collinear spin structure,[20,21] but an interfacial DMI arises due to the coexistence of (i) strong spin-orbit coupling in $Bi_2Se_3$ and (ii) broken inversion symmetry of the $Bi_2Se_3/BaFe_{12}O_{19}$ bi-layer. This DMI competes with exchange interaction and, under certain conditions, can produce skyrmions in $BaFe_{12}O_{19}$. As electrons in $Bi_2Se_3$ move along the surface at the interface and across through regions right above the skyrmions, they interact with the magnetic moments in the skyrmions and thereby experience a fictitious magnetic field ($b$) from them. This fictitious field is associated with the real-space Berry phase of the skyrmions.[12,13,22] It deflects the electrons from their otherwise straight course and results in a Hall voltage, just as a magnetic field deflects electrons in the OHE. The interaction of the conduction electrons in $Bi_2Se_3$ with the magnetic moments in $BaFe_{12}O_{19}$ may be achieved through spin-dependent scattering at the interface (a non-equilibrium proximity effect),[23] or direct coupling to the moments induced in the $Bi_2Se_3$ atomic layers near the interface (an equilibrium proximity effect),[4,5,7,24,25,26] or both.

To support the above interpretation, it is necessary to examine the temperature and field dependences of the THE. In particular, four distinct features are expected. First, the THE should be present only in a certain field range. The formation of skyrmions involves a fine balance between different terms in the magnetic free energy, thus usually requiring an external perpendicular field of appropriate strength: if the field is too weak, one generally has a helical state; if too strong, one has a usual ferromagnetic state.[27,28]

Second, the THE strength should not be affected by a moderate in-plane field, because skyrmions, once formed, are topologically protected and should be robust against in-plane fields.

Third, the THE should be absent above a threshold temperature ($T_{th}$). This expectation results from the $T$ dependence of the DMI constant $D$. In general, the formation of highly packed skyrmions requires[27,28]

$$D \geq \frac{4}{\pi}\sqrt{AK} \quad (3)$$

where $A$ is the exchange constant and $K$ is the perpendicular anisotropy constant. Consider first that Eq. (3) is satisfied at low $T$. With an increase in $T$, both $D$ and $K$ decrease, but $D$ can decrease more significantly than $K$ and Eq. (3) can be violated at a certain temperature, resulting in the disappearance of the THE. Previous experiments found $D \propto M_s^{4.9}$ and $K \propto M_s^{2.1}$ for Pt/Co bi-layers, where $M_s$ decreases with $T$ according to Bloch's $T^{3/2}$ law.[29]

Fourth, the THE strength is expected to decrease when $T$ is increased towards $T_{th}$. This is because $\rho_{THE}$ can be evaluated as[12,13,22]

$$\rho_{THE} = \frac{P}{en_{2D}}b = \frac{P}{en_{2D}}\frac{\Phi_0}{d^2} \quad (4)$$

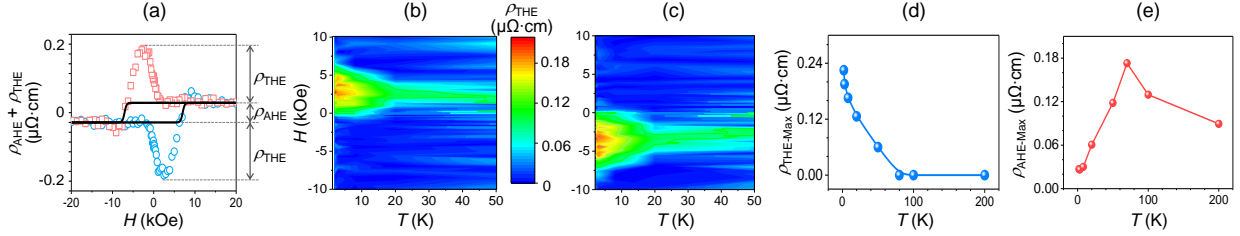

**Figure 3.** THE and AHE properties of Bi$_2$Se$_3$(6 nm)/BaFe$_{12}$O$_{19}$(5 nm). (a) Diagram illustrating how $\rho_{THE}$ and $\rho_{AHE}$ are separated. The black curves show fits to Eq. (5). (b) and (c) color maps showing $\rho_{THE}$ (color) as a function of $T$ and a perpendicular field ($H$). The field was swept up for (b) and down for (c). (d) $\rho_{THE\text{-Max}}$ vs. $T$. (e) Maximum $\rho_{AHE}$ vs. $T$.

where $P$ is the spin polarization rate in Bi$_2$Se$_3$, $e$ is the electron charge, $n_{2D}$ is the sheet carrier density in Bi$_2$Se$_3$, $\Phi_0$ is the magnetic flux quantum, and $d$ is the average distance between the centers of two neighboring skyrmions and scales with $\frac{2\pi A}{D}$.[27] $\rho_{THE}$ decreases with an increase in $T$ because an increase in $T$ leads to (i) a decrease in $D$ and a corresponding increase in $d$, and (ii) an increase in $n_{2D}$.[6]

To examine those expected features, control measurements were conducted from which the main results are presented in Figs. 3 and 4. Figures 3(b) and 3(c) give $\rho_{THE}$ as a function of $T$ and $H$ (perpendicular), where $\rho_{THE}$ is separated from $\rho_{AHE}$ by assuming[16]

$$\rho_{AHE}(H) = \rho_{AHE-Max} \tanh\left(\frac{H \pm H_c}{H_0}\right) \quad (5)$$

for a given $T$, as illustrated in Fig. 3(a). In Eq. (5), $H_c$ is the coercivity, while $\rho_{AHE\text{-Max}}$ and $H_0$ are fitting parameters. The red-yellow pockets in Figs. 3(b) and 3(c) indicate that the strongest THE occurs under $H\approx$2-5 kOe over $T$=2-10 K. Over $T$=30-50 K, the THE exists in a narrower field range, from ~2.5 kOe to ~5 kOe. These results are consistent with the first feature discussed above.

Figure 3(d) presents $\rho_{THE\text{-Max}}$ over $T$=2-200 K. The data, together with the maps in Figs. 3(b) and 3(c), show that there exists $T_{th}$ for the onset of the THE, which is ~80 K; as $T$ decreases from 80 to 2 K, the THE strength evidently increases. These results are the same as the third and fourth features discussed earlier.

Figure 4(a) shows the data measured at different field angles ($\theta$). Figures 4(b) and 4(c) present $\rho_{THE\text{-Max}}$ and the perpendicular component of $H$ ($H_\perp$=$H\cos\theta$) at which $\rho_{THE\text{-Max}}$ occurs, as a function of the in-plane component of $H$ ($H_\parallel$=$H\sin\theta$). One can see that an increase in $H_\parallel$ leads to a notable decrease in $H_\perp$, but only a minor change in $\rho_{THE\text{-Max}}$. The former result is consistent with the fact that skyrmion formation requires a fine balance between different energies, and a change in one energy should be accompanied by a change in another; the latter indicates the robustness of the skyrmions, which is the second feature expected.

Thus, the four expected THE signatures have all been confirmed. This fact, together with the analyses in Fig. 2, evidently establishes the observed THE as a genuine THE. To further support this, theoretical calculations and simulations were performed to examine the formation and properties of skyrmions in Bi$_2$Se$_3$/BaFe$_{12}$O$_{19}$. The main results are presented in Fig. 5, while more details are given in Supplemental Materials S6 and S7.

Figure 5(a) shows a phase diagram calculated by using experimental saturation magnetization ($M_s$) and anisotropy constant ($K$) values [see Supplemental

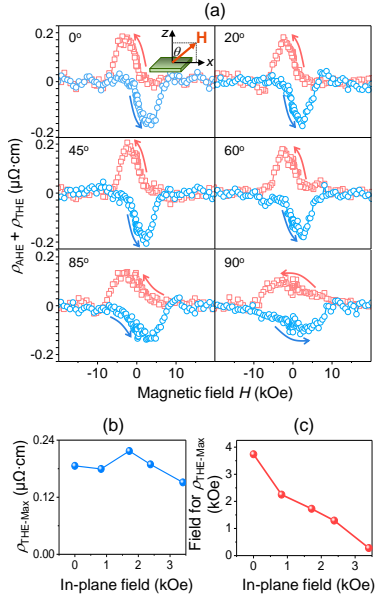

**Figure 4.** Effects of in-plane fields on the THE of Bi$_2$Se$_3$(6 nm)/BaFe$_{12}$O$_{19}$(5 nm). (a) $\rho_{AHE}$+$\rho_{THE}$ measured as a function of a field ($H$) for different field angles ($\theta$). (b) and (c) show $\rho_{THE\text{-Max}}$ and the perpendicular field as a function of the in-plane field.

Material S1] and taking into account the temperature dependences of $A$[30] and $D$[29] as

$$A(T) = A_0 \left(\frac{M_s}{M_{s0}}\right)^{\frac{5}{3}} = A_0 \left[1 - \left(\frac{T}{T_c}\right)^{\frac{3}{2}}\right]^{\frac{5}{3}} \quad (6)$$

$$D(T) = D_0 \left(\frac{M_s}{M_{s0}}\right)^{5} = D_0 \left[1 - \left(\frac{T}{T_c}\right)^{\frac{3}{2}}\right]^{5} \quad (7)$$

where $A_0 = 1.0 \times 10^{-11}$ J/m, [31,32] $D_0 = 3.0 \times 10^{-3}$ J/m², and $M_{s0} = 4.2 \times 10^5$ A/m are the values of $A$, $D$, and $M_s$ at $T=0$ K, respectively, and $T_c = 767$ K is the Curie temperature of $BaFe_{12}O_{19}$ [Supplemental Material S1]. The calculations utilized a Ginzburg-Landau model.[33,34,35] The $D_0$ value is chosen to let the $T$ window for the calculated skyrmion-lattice phase roughly agree with that for the measured THE. The main features of the calculated diagram are similar to those shown in Fig. 3(b); this similarity supports the above interpretation of the THE origin. Further, since $D_0$ is the only free parameter, one may expect that the actual $D_0$ value for $Bi_2Se_3/BaFe_{12}O_{19}$ is close to $3.0 \times 10^{-3}$ J/m².

Figures 5(b), 5(c), and 5(d) present the results simulated with $T=2$ K and $D_0=2.5 \times 10^{-3}$ J/m²; the other parameters in the simulations are the same as those used in the phase diagram calculations. Figures 5(b) and 5(c) give the magnetic morphology images simulated for $H=0$ and $H=2.5$ kOe, respectively. The colors denote the strength of the perpendicular component ($m_z$) of the normalized magnetization (**m**). Figure 5(d) compares

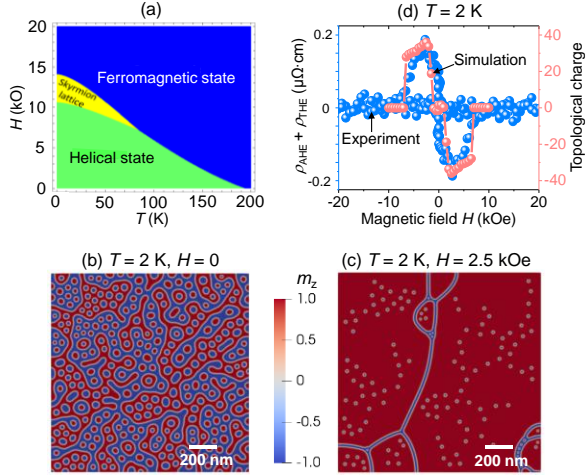

**Figure 5.** Theoretical and numerical results for THE and skyrmions in $Bi_2Se_3/BaFe_{12}O_{19}$. (a) A theoretical magnetic phase diagram. (b) and (c) show real-space magnetic morphology images obtained via simulations for different fields ($H$). (d) Comparison of experimental THE data and topological charge results from simulations.

the THE responses measured at $T=2$ K with the topological charge ($Q$) vs. $H$ response, where $Q$ was determined by

$$Q = \frac{1}{4\pi} \int \mathbf{m} \cdot \left(\frac{\partial \mathbf{m}}{\partial x} \times \frac{\partial \mathbf{m}}{\partial y}\right) dx dy \quad (8)$$

The image in Fig. 5(b) was obtained by initializing the magnetization in a random state at $T=2$ K and $H=0$ first and then relax. It shows random distribution of "positive" skyrmion bubbles with $m_z>0$ cores in the blue domains and "negative" skyrmion bubbles with $m_z<0$ cores in the red domains. The bubbles have an average size of ~50 nm. The numbers of two types of skyrmion bubbles approximately equal to each other, giving rise to $Q\approx0$. As $H$ is increased, the red domains expand, the blue domains shrink, the skyrmion bubbles with flat out-of-plane cores evolve into small skyrmions with point cores, having a size of ~20 nm, as shown in Fig. 5(c). At the same time, the shrinking of the blue domains results in the annihilation of "positive" skyrmions, leading to a change in $Q$ from 0 to a negative value. As $H$ is further increased, $|Q|$ decreases slightly due to the shrinking of some red domains, such as the one on the low-left corner in Fig. 5(c) [see Fig. S4(e)]. If $H$ is further increased to 7.5 kOe or higher, it saturates the $BaFe_{12}O_{19}$ film and annihilates all remaining skyrmions, which makes $Q$ change back to 0, as shown in Fig. 5(d). The nice agreement between the experimental and numerical responses shown in Fig. 5(d) supports the above interpretation of the observed THE.

It is important to emphasize that the path taken by the system in the simulations [Figs. 5(b)–5(d)] is along with local thermal equilibrium states, while the calculation of the phase diagram in Fig. 5(a) is for the system in global equilibrium states. The path of the experimental system does not proceed along with the global equilibrium states; if it did, there would be no hysteresis. Instead, the experimental system proceeds along a path that, as $H$ is increased from a strong negative field to zero, goes through local equilibrium states with $Q=0$ and then leads to a zero-field state with both "positive" and "negative" skyrmion bubbles nucleated due to defects or inhomogeneities in the sample. In other words, the experimental path is close to that in the simulations but differs from that in the phase diagram calculations. It is most likely for this reason that the field range of the skyrmions in the simulations agrees with that of the experimental THE region, but the fields for the closely-packed skyrmion lattice region in the calculations are higher.

It is for the same reason that the skyrmions in $BaFe_{12}O_{19}$ are likely to be randomly distributed, rather than being hexagonally packed, and have a size of ~20 nm. It is also believed that the skyrmions are uniform across the entire thickness of the $BaFe_{12}O_{19}$ thin film (5 nm) because of strong exchange interaction in $BaFe_{12}O_{19}$. It will be of great interest to see future work that directly probes spin textures in skyrmions in $Bi_2Se_3/BaFe_{12}O_{19}$ via Lorentz transmission electron microscopy,[36] magnetic force microscopy,[37] or scanning transmission X-ray microscopy[42] measurements, although it may be challenging as the measurements need to be done at low temperatures and $BaFe_{12}O_{19}$ is insulating.

Two notes should be made. First, $T_{th}$ in Fig. 3 is ~80 K, but THE responses in other MIs can occur at much higher $T$.[38,39] This is because $K$ in $BaFe_{12}O_{19}$ is significantly higher than in other MIs; one can extend $T_{th}$ to room temperature by reducing $K$ in $BaFe_{12}O_{19}$ via doping.[40] Second, Figure 3(e) shows that as $T$ is varied from 80 K to 2 K, $\rho_{AHE-Max}$ decreases significantly. This response results from the $T$ dependence of the conductivity of the TSS and the phonon density (Supplemental Material S8).[6]

In closing, two remarks should be made about the DMI strength. First, $D_0$ used in the simulations ($2.5\times10^{-3}$ J/m$^2$) is smaller than that in the phase diagram calculations ($3.0\times10^{-3}$ J/m$^2$); they were chosen to let $T_{th}$ from the simulations and the calculations roughly match $T_{th}$ measured. The use of two different $D_0$ values is not inappropriate because the paths taken by the system in the simulations and the calculations are different, as discussed above. Second, since the DMI strength dictates the skyrmion size, it is interesting to compare the DMI strength in $Bi_2Se_3/BaFe_{12}O_{19}$ with that reported previously. Considering the interfacial nature of the DMI and the bulk nature of the skyrmions (uniform across the $BaFe_{12}O_{19}$ thickness), one can evaluate the DMI by defining a DMI energy parameter $E_{DMI} = Dt$ ($t$: MI thickness).[41] Based on previously reported $D$ and $t$, one obtains $E_{DMI} \approx 1.17$ pJ/m for Pt/Co/Ta,[42] 0.96 pJ/m for Ir/Co/Pt,[43] 2.17 pJ/m for Pt/Co/MgO,[41] 0.44 pJ/m for W/CoFeB/Mgo,[44] and 0.02 pJ/m for Pt/Tm$_3$Fe$_5$O$_{12}$.[15] In contrast, one has $E_{DMI} \approx 12.5$ pJ/m in this work. Thus, the DMI strength in $Bi_2Se_3/BaFe_{12}O_{19}$ is one or two orders of magnitude higher than that in heavy metal-based systems. Such a strong effect can be attributed to the spin-momentum locking in $Bi_2Se_3$.

*Contributed equally; †mwu@colostate.edu

**Acknowledgement:** See Ref. [45].

**Supplemental Material** is available free of charge via the internet. It includes the details about material growth and characterization, Hall bar devices, anti-symmetrization and numerical fitting of Hall resistance data, calculations, and micromagnetic simulations.

[45] Acknowledgement: This work was supported by the U.S. Department of Energy, Office of Science, Basic Energy Sciences (DE-SC0018994). The fabrication and characterization of the measurement samples were supported by the U.S. National Science Foundation (EFMA-1641989; ECCS-1915849). Instrumentation supported by the National Science Foundation MRI program (DMR-1727044) was used for this work. Work at Argonne National Laboratory was supported by the U.S. Department of Energy, Office of Science, Basic Energy Sciences Division of Materials Sciences and Engineering. Work at CWRU was supported by the College of Arts and Sciences at CWRU. Work at PSU was supported by the Penn State Two-Dimensional Crystal Consortium-Materials Innovation Platform (2DCC-MIP) under the U.S. National Science Foundation Grant No. DMR-1539916. Work at UW was supported by the U.S. National Science Foundation (DMR-1710512) and the U.S. Department of Energy, Office of Science, Basic Energy Sciences (DE-SC0020074). Work at UA is supported by the U.S. National Science Foundation (ECCS-1554011). The authors acknowledge Dr. Vijaysankar Kalappattil and Mr. Yuejie Zhang for helping with transport measurements and Mr. Laith Alahmed for helping with data analyses.